# DEVELOPMENT OF THE 50-MEV DTL FOR THE JAERI/KEK JOINT PROJECT

F. Naito, K. Yoshino, C. Kubota, T. Kato, Y. Saito, E. Takasaki, Y. Yamazaki, KEK,
1-1 Oho, Tsukuba-shi, Ibaraki-ken,305-0801 Japan
S. Kobayashi, K. Sekikawa, M. Shibusawa, Saitama University
Shimo-Okubo, Urawa, 338-8570 Japan
Z. Kabeya, K. Tajiri, T. Kawasumi, Mitsubishi Heavy Industry
10 Oye-cho, Minato-ku, Nagoya, 455 Japan

*Abstract:*

An Alvaretz-type DTL, to accelerate the $H^-$ ion beam from 3 to 50 MeV, is being constructed as the injector for the JAERI/KEK Joint Project. The following components of the DTL have been developed: (1) a cylindrical tank, made by copper electroforming; (2) rf contactors; (3) a pulse-excited quadrupole magnet with a hollow coil made by copper electroforming; (4) a switching-regulator-type pulsed-power supply for the quadrupole magnet. High-power tests of the components have been conducted using a short-model tank. Moreover a breakdown experiment of the copper electrodes has been carried out in order to study the properties of several kinds of copper materials.

## 1. INTRODUCTION

Construction has started of an Alvaretz-type DTL, to accelerates the $H^-$ ions from 3 to 50 MeV, as the injector as part of the JAERI/KEK Joint Project at the high-intensity proton accelerator facility in Japan. The DTL consists of three long tanks (maximum 9.9 m in length), each of which is comprised of three short unit tanks (approx. 3 m in length), to overcome difficulties with constructing the tank and assembling the drift tube. The resonance frequency of the DTL is 324 MHz. The rf pulse length is 600 μsec and its repetition rate is 50 Hz. The main design parameters of the DTL are summarized in Table-1 [1]. This report describes three aspects of the DTL construction; rf contactors, the characteristics of the copper surface of the cavity, and the power supply for the quadrupole magnet in the drift tube.

Table 1. DTL design parameters

| Tank | No. 1 | No. 2 | No. 3 |
|---|---|---|---|
| Energy (MeV) | 19.7 | 36.7 | 50.1 |
| No. of Cell | 76 | 43 | 27 |
| Length (m) | 9.92 | 9.44 | 7.32 |
| Tank dia. (mm) | 561.1 | 561.1 | 561.1 |
| DT dia. (mm) | 140 | 140 | 140 |
| Stem dia. (mm) | 34 | 34 | 34 |
| Bore dia. (mm) | 13, 18 | 22 | 26 |

## 2. RF CONTACTOR

Two types of rf contactor have been developed: (a) a contactor between the end plate and the tank cylinder; and (b) a contactor between the stem of the drift tube and the tank. Cross-sectional views of the contactors are shown in Figure 1. The structure is very simple: a thin copper layer (0.5 mm in thickness) surrounds a stainless steel spring. There is a vacuum seal outside the rf contactor. The performance of these was checked initially by a small test cavity and then by the large cavities that are described in the next section.

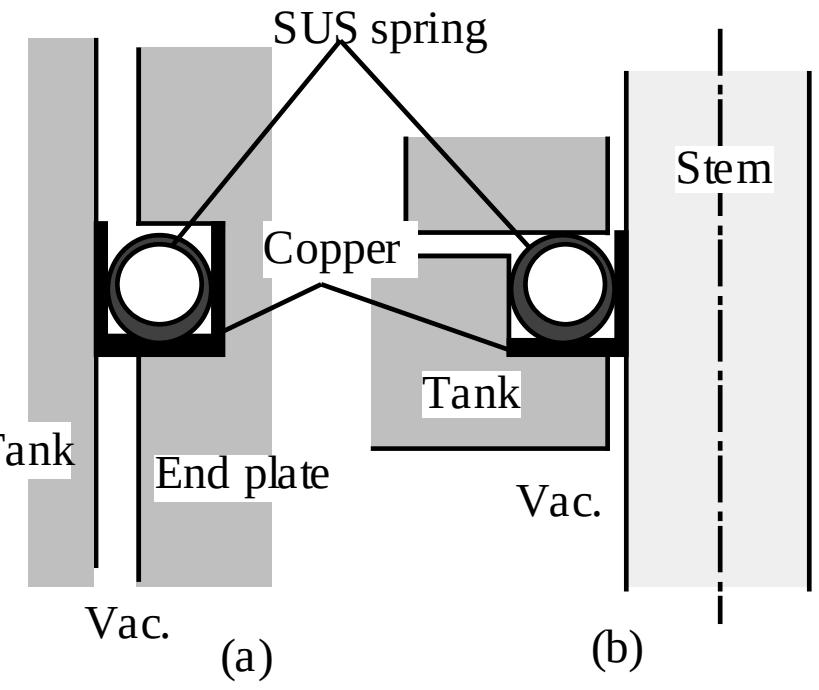

Figure 1. rf-contactors
(a) rf-contactor between the end-plate and the tank, (b) rf-contactor between the stem of the drift tube and the tank

## 3. VACUUM AND RF PROPERTIES OF THE TEST TANK

The cavity cylinder for the DTL is made of iron, with the inner surface covered by a copper layer (0.5 mm in thickness) that was built by the Periodic Reverse (PR) electroforming using pure copper sulphate bath, and then finished by electropolishing [2]. A cylindrical cavity was made to test the vacuum and the rf properties of the PR copper electroforming surface. The size of the cylinder (560 mm in diameter, 3320 mm in length) is almost identical to that of the longest unit tank of the DTL. The rf contactor described in the previous section is used for the end plates.

The measured unloaded Q-value of the $TM_{010}$ mode of the cavity is 77000, which represents approximately 97 % of the value obtained by analytical calculations. The results indicate that (a) the electrical quality of the copper surface is sufficiently high and (b) that the rf contactor functions satisfactorily at a low rf-power level.

Vacuum property was also measured, and as the results in Figure 2 show, the pressure level of the tank became $10^{-5}$ Pa after 100 hours of evacuation. The outgas rate from the tank surface was also measured by an integration method (build-up test) and the data is presented in Figure 3. The outgas rate for the 2nd measurement was $5.2x10^{-8}$ Pa $m^3/s/m^2$ ($3.9x10^{-11}$ Torr $l/s/cm^2$), which is closely consistent with the value for the OFC.

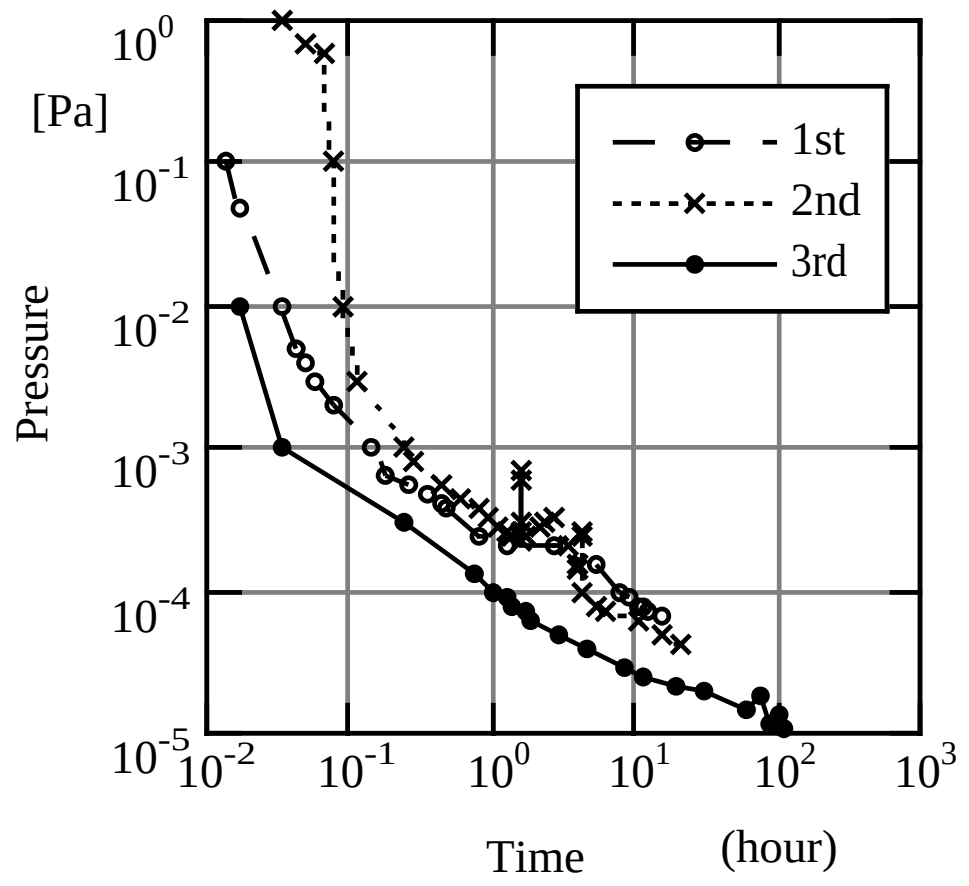

Figure 2. Ultimate pressure for the 3m tank

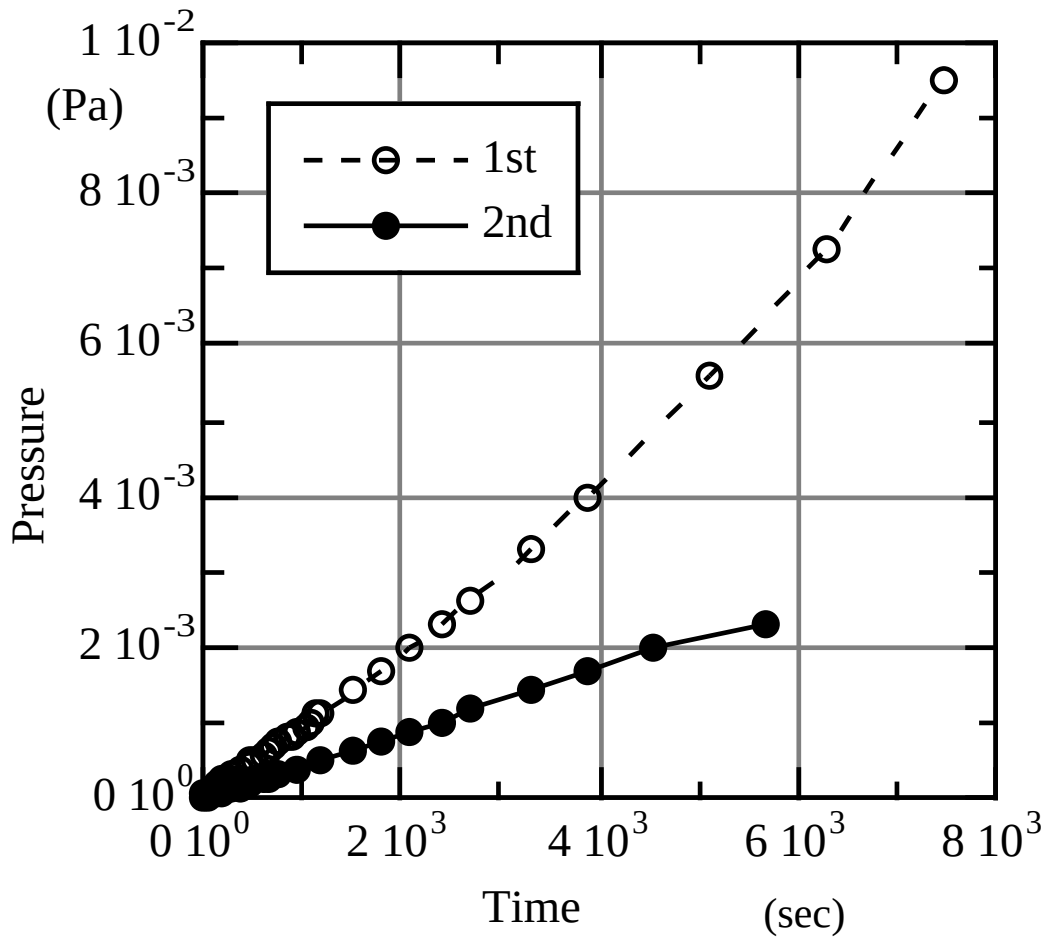

Figure 3. Pressure variation in the tank with outgas.

## 4. QUADRUPOLE MAGNET AND PULSED-CURRENT SUPPLY

One of the most important devices for the DTL is the quadrupole magnet in the drift tube. We have developed a compact quadrupole electromagnet with a hollow coil made by the PR copper electroforming. Because the magnet is operated in pulse mode to decrease the heating due to ohmic loss in the coil, the pole piece of the magnet is made from a stack of silicon steel plate (0.35 mm and 0.5 mm in thickness). Details of the magnet are reported in reference [3].

A pulsed-current supply with a 20 kHz switching regulator circuit (IGBT elements are used) has been developed for the magnet. The requirements for the current supply are as follows: (1) current stability, with the flat top of the output pulse being less than $10^{-3}$; (2) the duration time of the stabilized flat top should be greater than 1 msec; (3) the maximum current is 1000A; (4) the rise time for the current pulse is 5 msec. A typical measured pattern of the output current from the supply with a dummy coil is shown in Figure 4. This shows that the stability of the current is about $5x10^{-4}$. The other requirements have also been achieved.

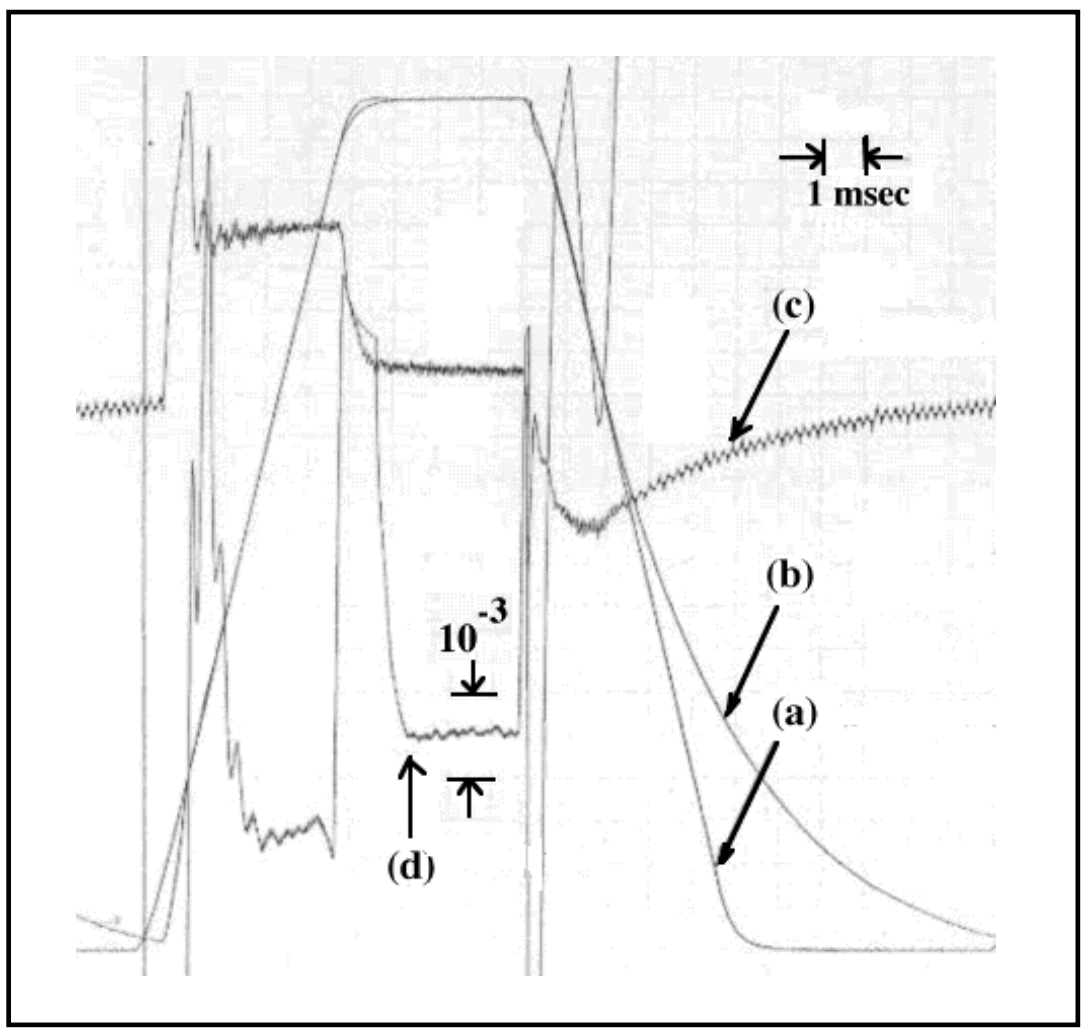

Figure 4. Time variation in the current from the power supply. (1000A)
(a) reference pulse (input), (b) output current,
(c) output voltage, (d) deviation of the output current

## 5. BREAKDOWN TEST OF THE COPPER ELECTRODE

An electrical breakdown test has been conducted to determine the electrical characteristics of the electroformed copper by the PR process for the DTL [4,5]. Electrodes made by other processes were also tested, in order to compare their properties. The top of the electrode is hemispheric in shape, with a radius of 30mm.

The results show that the first breakdown field for the electrode made by the PR copper electroforming is significantly higher than those of the other electrodes, except for an electrode made of OFC finished by a diamond bite. Table 2 shows typical results for the first breakdown field level. The data

Table-2. The first breakdown field

| Materials | The 1st breakdown field (MV/m) |
|---|---|
| EF (PR, Pure copper sulfate) | 41 |
| EF (Copper sulfate with brightener) | 13 |
| EF (Pyrophosphate) | 10 |
| OFC (Lathe finishing) | 20 |
| OFC (Electro polishing) | 16 |
| OFC (Diamond bite) | 70 |

(EF: Electro-Forming, PR: Periodic-Reverse
OFC: Oxygen Free Cooper)

indicates that the surface of the electroformed copper by PR process has the most suitable properties for the accelerator cavity.

## 6. HIGH-POWER TEST OF A MODEL TANK

A short tank (1.4 m in length) has been made for a high-power test of the DTL components ( the rf contactors for the stems and the end plates, the electroformed copper surface, a drift tube with the quadrupole magnet, tuners, and the input coupler). A schematic representation of the tank is shown in Figure 5. Only the shortest drift tube has a quadrupole magnet inside. The tank consists of the first three cells and the last four cells of the DTL. The right half of the 3rd drift tube is half of the 142nd drift tube. Thus, the position of the stem of the tube is not ideal, as the stem is located at center of the tube. As this is likely to lead to non-uniformity in the accelerating field (Ez) distribution, which is measured by a bead perturbation method, the non-uniformity around the 3rd drift tube, that can be seen in the data shown in Figure 5, is as expected. The measured unloaded Q-value was 46200, which is about 93 % of the estimated value, and includes the effect of all components. The shunt impedance is 54.7 MΩ/m. Because the design value of the Ez is 2.5 MV/m, the required input rf-power is about 160 kW.

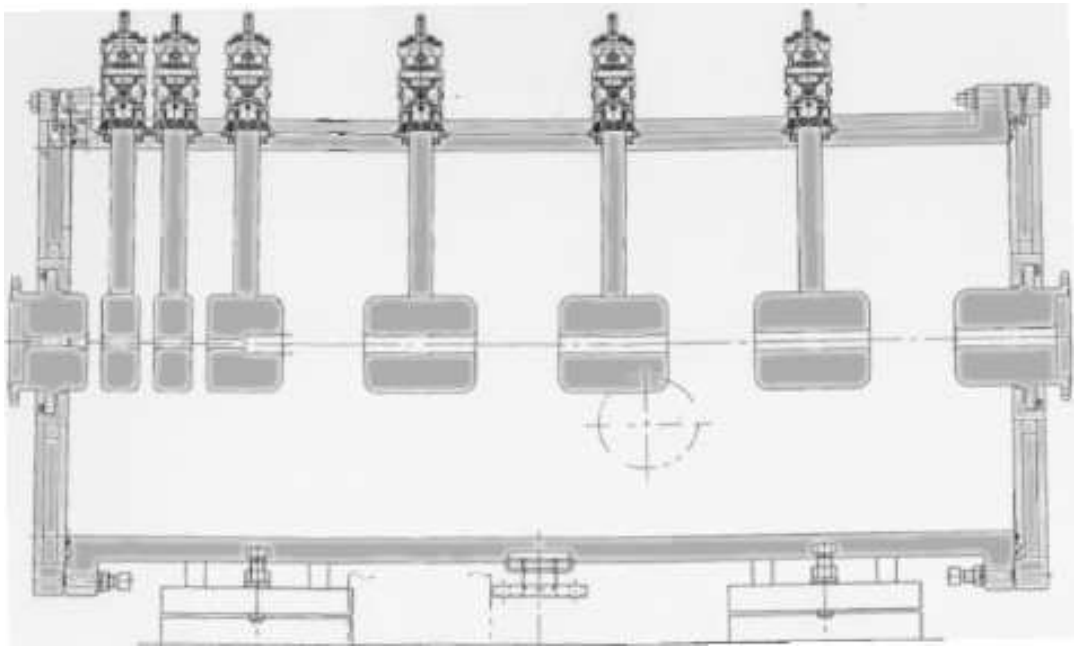

Figure 5. Schematic representation of the DTL hot model

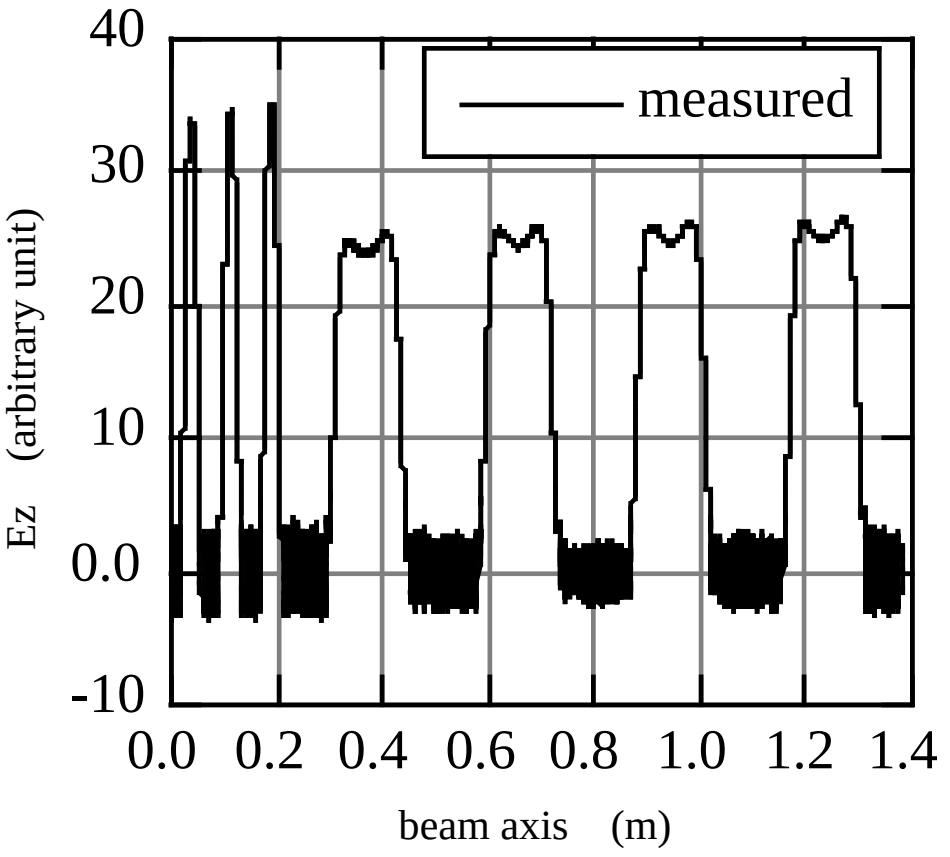

Figure 6. Electric field along beam axis

The first high-power conditioning was carried out at the end of April this year. Figure 7 shows the conditioning history of the tank. The design value for the peak power was easily achieved with the short-pulse (several 10 μsec in duration) operation, requiring just two days to achieve full-power operation. The high-power test was terminated when a ceramic window of the input coupler broke; however, when the end plate of the tank was opened to check the inside, no trace of the discharge was observed on the inner surface of the tank ,the surface of the drift tube, or on the rf contactor for the end plate.

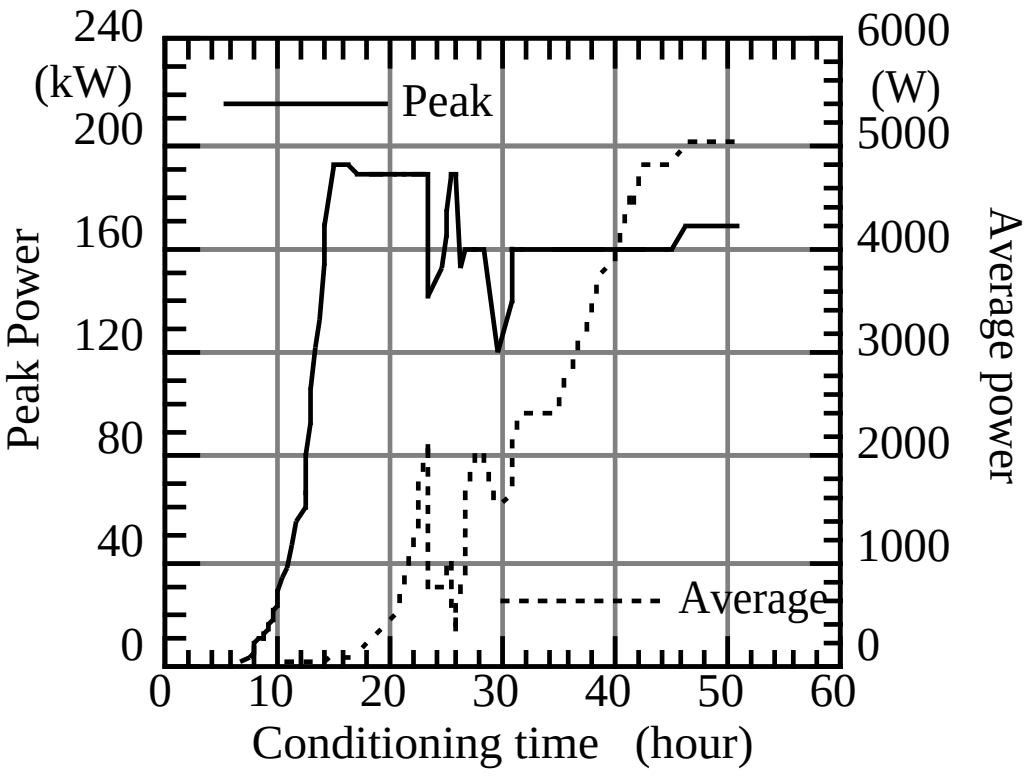

Figure 7. Conditioning history of the model

## 7. CONCLUSION

The construction of the Alvaretz DTL for the JAERI/KEK joint project has been started. The components developed for the DTL have been tested in a high-power test of the DTL model. In particular, the PR electroformed copper was found to have excellent properties. There are problems with the input coupler that need to be solved. However, mass production of the magnet and drift tube is already in progress, and the production of the pulsed-power supplies for the first DTL tank is completed.